\documentclass[10pt,conference]{IEEEtran}

\usepackage{amsmath,amsfonts,amssymb,graphicx}
\usepackage{algorithm,algpseudocode}
\usepackage{caption}
\usepackage{subcaption}
\usepackage[mathscr]{eucal}

\usepackage{mathtools}
\usepackage[subnum]{cases}

\usepackage{float}

\usepackage{url}

\usepackage{verbatim}

\usepackage{booktabs}
\usepackage{multirow} 

\usepackage{fixltx2e}

\newcommand{\Abs}[1]{\ensuremath{\left\vert #1 \right\vert}}

\newcommand{\TwoNorm}[1]{\ensuremath{ \Norm{#1}_{2}}}

\newcommand{\FNorm}[1]{\ensuremath{ \Norm{#1}_{F}}}
\newcommand{\Norm}[1]{\ensuremath{ \left\Vert #1 \right\Vert}}

\newcommand{\Mat}[1]{\ensuremath{ \boldsymbol{\mathbf{#1}}}}
\newcommand{\Vect}[1]{\ensuremath{ \boldsymbol{\mathbf{#1}}}}

\newcommand{\R}{\ensuremath{\mathbb{R}}}
\newcommand{\Rn}{\ensuremath{\R^n}}
\newcommand{\argminy}{\ensuremath{\displaystyle\operatornamewithlimits{arg\,min}_y}}
\newcommand{\thinColon}{%
  \nobreak
  \mskip1mu 
  \mathpunct{}%
  \nonscript
  \mkern-\thinmuskip
  {:}%
  \mskip1mu
  \relax
}
\newcommand{\IndexRange}[2]{\ensuremath{{#1}\thinColon{}{#2}}}

\title{Evaluating the Impact of SDC on the GMRES Iterative Solver}

\author{
\IEEEauthorblockN{
James Elliott\IEEEauthorrefmark{1}\IEEEauthorrefmark{2},
Mark Hoemmen\IEEEauthorrefmark{2}, and
Frank Mueller\IEEEauthorrefmark{1}}
\IEEEauthorblockA{
\IEEEauthorrefmark{1}
Computer Science Department, North Carolina State University,
Raleigh, NC
}
\IEEEauthorblockA{
\IEEEauthorrefmark{2}
Sandia National Laboratories
Albuquerque, NM
}
}

\begin{document}

\maketitle

\begin{abstract}
Increasing parallelism and transistor density, along with increasingly tighter
energy and peak power constraints, may force exposure of occasionally incorrect
computation or storage to application codes. Silent data corruption (SDC) will
likely be infrequent, yet one SDC suffices to make numerical algorithms like
iterative linear solvers cease progress towards the correct answer. Thus, we
focus on resilience of the iterative linear solver GMRES to a single transient
SDC.
We derive inexpensive checks to detect the effects of an SDC in GMRES that
work for a more general SDC model than presuming a bit flip.
Our experiments show that when GMRES is used as the inner solver of an
inner-outer iteration, it can ``run through'' SDC of almost any magnitude in the
computationally intensive orthogonalization phase.
That is, it gets the right answer using faulty data without any required roll
back. Those SDCs which it cannot run through, get caught by our detection
scheme.

\end{abstract}

\section{Introduction} \label{jje:sec:intro}
Incorrect arithmetic or corruption of stored data could have dire
effects on the execution of a numerical algorithm.  Experiments show
that a single bit flip in memory can cause certain algorithms to
``crash'' (terminate abnormally, due to invalid states or actions
detected by the application or operating system), ``stagnate'' (keep
running but fail to make progress), or, worst of all, produce the wrong
solution, silently.

Rather than focusing exclusively on bit flips, this work studies the
impact of Silent Data Corruption (SDC) on the Generalized Minimal
Residual Method (GMRES) iterative linear solver.  The source of the
corruption, while interesting, gives no insight into its impact on the
algorithm and the correctness of its result.  By generalizing bit
flips in floating-point data into potentially unbounded numerical
errors, we are able to use mathematical analysis both to reason about
algorithms' behavior should an SDC event occur, and to harden them
against the event's effects.

Fortunately, some numerical algorithms only need reliability for
certain data and phases of computation.  If the system can guard just
those parts of the algorithm in space and time, then the algorithm can
compute the right answer --- or at least be able to detect failure and
report it ``loudly'' --- despite faults in unreliable phases of
execution.  This suggests a ``layered'' approach to the design of
reliable numerical algorithms.  A reliable outer layer can recover
from faults in a less reliable inner layer.  If the solver spends most
of its time in unreliable mode, it can mitigate the cost of reliable
computation in the outer mode.
We begin the analysis with GMRES, and
then extend it to the Fault-Tolerant GMRES (FT-GMRES) inner-outer
iteration.

\textbf{We present the following contributions}:
\begin{itemize} 
\item We use mathematical analysis of the GMRES algorithm to
  construct a detector that bounds the error that SDC may introduce.
\item We combine the above detection scheme with the \emph{sandbox}
  reliability model presented in \cite{jje:bridges2012fault}.
\item We illustrate experimentally that bounded error originating in the faulty
inner solve has little impact on time-to-solution.
\end{itemize}

\subsection{Silent Data Corruption}

In this work, we address a very specific type of fault, i.e., a
fault that silently introduces bad data, while not persistently
tainting the data that was used in the calculation.  For
example, let $a=2$ and $b=2$, then $c = a + b = 10$, while simplistic,
this model presumes no knowledge of the nature of the fault, only that
$c$ is incorrect.  This model assumes that the machine is unreliable
in an unpredictable way, and therefore we are skeptical of the output
it presents.  This type of unreliability can be mitigated via
redundant computation and introspection, but then the cost of running
the algorithm increases drastically.

\subsection{Faults, Failures, and Persistence}

\begin{figure}
    \centering
    \includegraphics[width=\columnwidth]{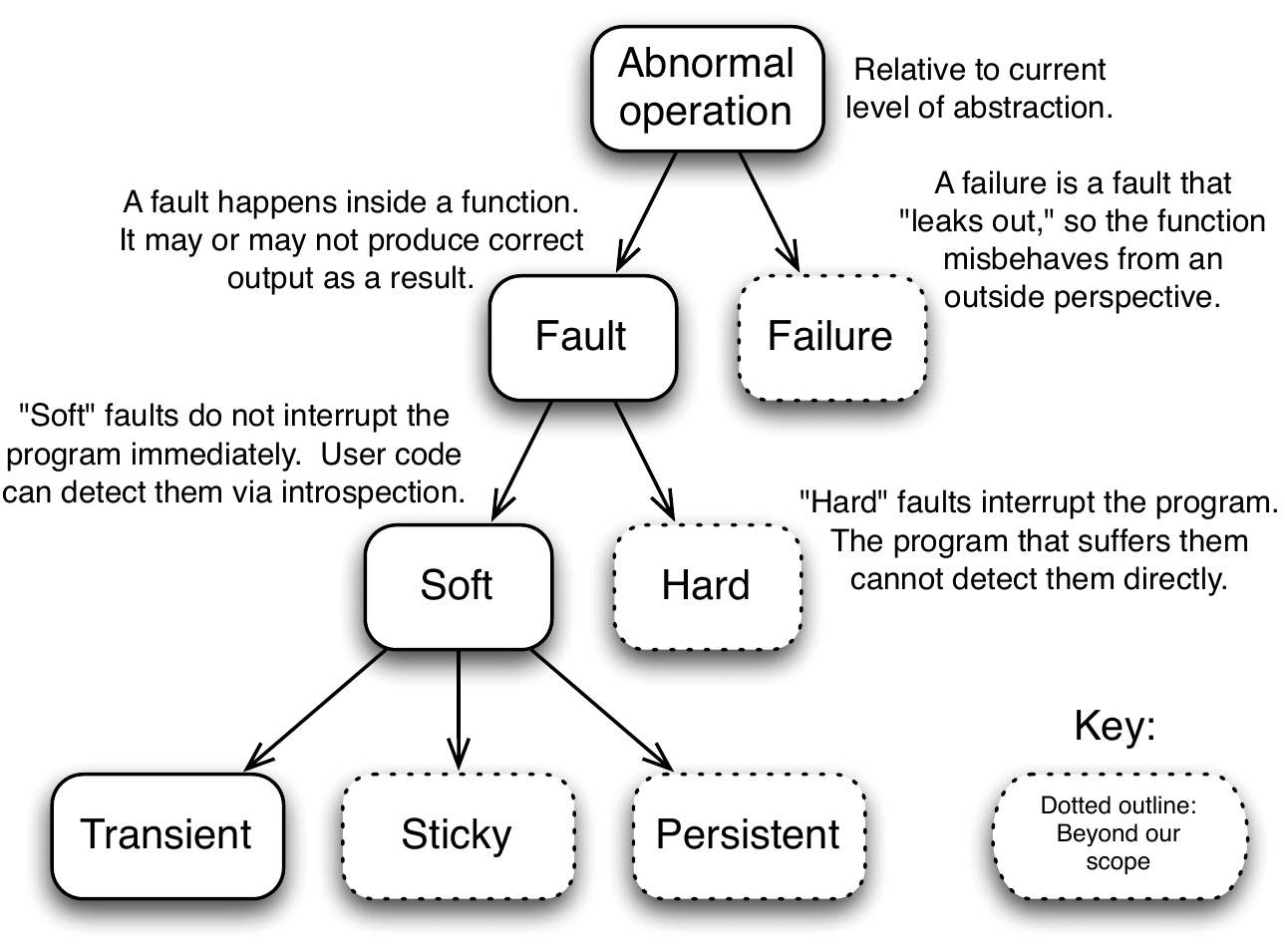}
    \caption[labelInTOC]{Taxonomy of faults and scope of this work.}
    \label{jje:fig:FaultTree}
\end{figure}

Our goal is to ensure that should transient SDC occur, we either
obtain the correct solution or make the fault not silent by alerting
the user.  We consider two perspectives: the user and the system.  A
fault occurs at the system level, e.g., a bit flips or a node
crashes. \emph{A fault becomes a failure if it impacts the user.}
Figure~\ref{jje:fig:FaultTree} depicts a visual taxonomy of how we
consider faults and the scope of our work.  We further classify faults
into those that interrupt the user's program (hard faults), and those
that do not immediately or ever interrupt the user's code (soft
faults).  A hard fault results in a failure if the user is running an
application (though a checkpoint / restart recovery system can ``mask
out'' hard faults, making them not failures).  In contrast, the very
nature of soft faults implies that they may emit no indication that
something has gone wrong.  In the event that soft faults allow the
program to continue execution with tainted data, we must understand
how algorithms behave in the presence of faulty data.  Furthermore, if
the algorithm uses tainted data and still obtains the correct
solution, then the fault does not constitute a failure.  If the soft
fault leads to an incorrect solution, then the fault leads to a
\textbf{\textit{silent} failure}, which is an outcome we wish to make
very rare or impossible.

We further classify soft faults by how long the underlying hardware
remains faulty.  Persistent faults arise from hardware that is
permanently faulty, e.g., a stuck bit in memory, or the Intel Pentium
FDIV bug \cite{jje:intel:fdiv}.  Sticky faults indicate hardware that
is faulty for some duration but returns to normal operation.  Transient faults occur once, and while the fault is transient the
effect of the fault may be persistent.

\section{Project Overview} \label{jje:sec:project}

To quantify the possible effects of a single silent data corruption
event in GMRES, we propose a multifaceted approach.  We combine the
sandbox reliability model from \cite{jje:bridges2012fault}, Flexible
GMRES from Saad \cite{jje:saad1993flexible}, and mathematical analysis
of the GMRES algorithm to create a nested solver strategy that
combines an unreliable inner solver with a reliable outer solver,
while enforcing that, should SDC occur in the unreliable phase, the
error is bounded.  We then show through experiments how our scheme
``runs through'' single SDC events in the unreliable solver.
Using this approach, we ultimately seek to present analyses of
solvers such that we can quantitatively choose solvers based on their
resiliency to single events of SDC.
\\
\\
\emph{This paper is organized as follows}:
\begin{enumerate}
\item In Section~\ref{jje:sec:sandbox}, we describe the sandbox
  reliability model.
\item In Section~\ref{jje:sec:gmres}, we present standard GMRES, and
  uncover through mathematical analysis an invariant in the Arnoldi
  process contained within GMRES.
\item We describe how to check this invariant when using Modified
  Gram-Schmidt orthogonalization.
\item In Section~\ref{jje:sec:ftgmres}, we introduce Saad's Flexible
  GMRES and Bridges et al.'s FT-GMRES, and explain the relation between
  flexible solvers and the sandbox reliability model.
\item In Section~\ref{jje:sec:results}, we compose a nested solver
  using the Trilinos framework \cite{jje:Trilinos}, and present
  experimental results illustrating how our invariant check impacts
  time-to-solution.
\end{enumerate}
%

%
%

\subsection{Assumptions and Justification}
\label{jje:sec:intro:assumptions}
We restrict SDC to the numerical data used and generated by the
algorithm.  We explicitly exclude faults in control flow, data
structures, loop counters and other metadata used to implement the
algorithm.  The reason for this exclusion is that these issues
represent a different class of problems.

Our assumption that SDC occurs only once is fundamental. The implied
source of SDC is typically a bit flip, but we do not restrict our
model to silent bit upsets, given that there is limited data
available today to base such a model on. We justify our choice of single
transient SDC, based on what we do know about bit flips and the
reliability of the system:
\begin{enumerate}
  \item Hardware employs techniques to ensure that so-called ``single
    event upsets'' (SEUs) -- that is, bit flips -- do not occur.
    Therefore, it is expected that SEUs will be rare events.
  \label{jje:enum:assumptions:rare}
  \item If we can understand the best- and worst-case scenarios for
  the error that an SDC can contribute, we will have a baseline to
  conjecture about multiple bit flips, i.e., multiple occurrences of
  SDC.
  \label{jje:enum:assumptions:model}
  \item There currently is no solid theory, e.g., a statistical
  distribution, of the rate at which bit flips occur. Therefore,
  speculation about flip rates may or may not prove useful.
   \item Assuming a
  particular fault rate makes bold assertions about future hardware,
  especially given the reluctance of hardware manufacturers to divulge
  this information.
\end{enumerate}
By following this research path, we are able to avoid the
pitfalls presented in items 3 and 4 above, and we are able to isolate the impact
of SDC without other factors polluting our analysis.
\section{Motivation} \label{jje:sec:motivation}
Energy and peak power increasingly constrain modern computer
hardware, yet hardware approaches to protect computations and data
against errors cost energy. This holds at all scales of computation,
but especially for the largest parallel computers being built and
planned today. This results from a confluence of factors:
\begin{itemize}
\item Increasing parallelism (and therefore more components to fail)
  \cite{jje:asanovic2006landscape,jje:asanovic2009landscape}
\item Decreasing transistor feature sizes, making individual
  components more vulnerable
\item Extremely tight peak power requirements \cite{jje:kogge2008exascale},
  limiting the use of hardware redundancy to increase reliability
\end{itemize}
As these trends continue, hardware vendors may succumb to the
temptation to expose incorrect arithmetic or memory corruption to
application codes
\cite{jje:karnik2004characterization,jje:kogge2008exascale,jje:miskov-zivanov2007soft}.
Some studies already indicate that this behavior is appearing at the
user level \cite{jje:haque2010hard}.  In fact, some researchers
actively promote relaxing hardware correctness to save energy
\cite{lammers2010era}.

\subsection{Relation to Prior Work}
Much of the prior work on fault-tolerant iterative solvers has taken the
approach of assuming some fault model for bit flips, and then injecting bit
flips into specific numerical operations
\cite{jje:iterative:Shantharam:2011,jje:iterative:Shantharam:2012,jje:iterative:Sloan:2012},
or treating the application as a black box and injecting bit flips arbitrarily
\cite{jje:iterative:Bronevetsky:2008}.  A popular operation to analyze is
sparse matrix-vector multiply
\cite{jje:iterative:Shantharam:2011,jje:iterative:Sloan:2012}, a key
kernel in iterative linear solvers.  These approaches typically engineer a response
that mitigates, detects, or detects and corrects bit flips injected following
the assumed fault model. The focus in this type of research has been to detect
errors, and then respond -- e.g., correct the tainted
values, or roll back and resume computation from an assumed valid
state -- assuming that the fault does not occur frequently enough to
cause stagnation. 

In addition, all prior work on sparse iterative
methods is based on a fault model that assumes multiple bit flips
injected at some rate.  Most studies are also carried out with little care for
whether the bit flipped is a $0\to1$ or a $1\to0$, and most studies flip bits at
random locations.
We question many of the assumptions made, and in general question the
research approach.

\subsubsection{SDC is a rare event}

We begin by questioning models and experiments that assume SDC
happens at a sufficiently high rate for multiple events to occur in a
single linear solve.

We have a strong reason to believe that SDC is a rare event. Hardware
incorporates a fairly large amount of safeguards in-place to protect data and
instructions. For example, Intel provides the
Machine Check Architecture, which provides reporting of bit errors at the
register, cache (L1-L3), QuickPath Interconnect, and DRAM (via ECC) layers.
We do not attempt to conjecture about the likelihood of bit flips, rather we
turn to the theoretical basis that an algorithm is built on, and study how the
algorithm behaves when perturbed within the bounds imposed by
mathematical analysis. 

Current research by Michalak \cite{jje:Michalak:2012} found SDC
occurred rarely.
Michalak et al.\ placed a Roadrunner node in front of a neutron cannon
and bombarded it with particles~\cite{jje:Michalak:2012}.
  While the neutron fluxes are far beyond realistic, their
  observations showed a startlingly low occurrence of SDC, while
  outright node failure occurred far more frequently. Why then has
  current SDC research focused on failure rates?

  In practice, SDCs should remain rare, even at extreme scales of
  parallelism.
  Nevertheless, little if any current research has attempted to
  explain how a single SDC event impacts an algorithm and ultimately
  the solution. Research on how to counter multiple bit flips has not
  provided additional insight on the cause/effect relationship.

Also, an application may attribute much of its run time to linear solves, but
typically these are multiple linear solves, e.g., an implicit time stepping
algorithm that solves a nonlinear system at each time step. For example, see 
M{\"u}ller and Scheichl where a nonlinear system of size $10^{10}$ is
solved and the linear solver is restricted to $0.003$ seconds per solve
\cite{jje:Mueller2013preprint}.
  
\subsubsection{Fault Models and Silent Data Corruption}
  At a higher level, we challenge the research approach of assuming a fault
  model for SDC.
  By definition, the origin of silent data corruption is unknown,
  with one such origin being a silent bit flip.
    Instead of characterizing SDC, the studies propose solutions to a problem we
  understand only poorly. It is our goal first to analyze the effects of SDC, and
  then to propose both specific algorithmic techniques and general heuristics
  that minimize its impact, should it occur.  With this ability,
  mathematicians, scientists, and engineers can take quantifiable steps to
  develop algorithms and applications that are inherently resilient to SDC.
  
  In numerical algorithms using IEEE-754 floating-point data,
  regardless of the cause, SDC will produce either numeric values or the non-numeric infinity
  (Inf) and not-a-number (NaN) values. Injecting bit flips will produce either
  type of error, making the act of injecting a bit flip to study
  transient SDC unnecessary  as the
  outcome could have been achieved by merely setting the memory location equal
  to some value. We know from the IEEE-754 specification precisely what numeric
  values are possible, and given the mystery of how, when, and where SDC
  originates, any of the possible floating-point values are plausible.

  We advocate a drastically different approach,
  namely that SDC impacts the underlying mathematical assumptions that
  guarantee convergence of an algorithm. Rather than focusing on
  detecting binary errors, we treat bit flips as numerical errors and evaluate
  how these errors relate to the theoretical basis that the algorithm is built
  on. In this sense, we filter values that are theoretically impossible, while
  accepting variations that are allowable by the theory. While our approach does
  not ``solve'' the SDC problem, we exploit modern mathematical techniques,
  so-called flexible solves,  to cope with the bounded error we ``run through''.

\subsection{Invariants as Detectors}
Numerical algorithms often have invariants that they can check
inexpensively to decide whether hardware faults have corrupted an
intermediate result enough for it not to be useful. For example, Chen
\cite{jje:chen2013online} performs additional computation and parallel
communication in order to check invariants of the iterative linear
solvers GMRES \cite{jje:saad1986gmres}, CG, and BiCG. If those invariants
are violated, the solver can roll back one or more iterations and
resume from the last known correct point. In this work, we develop
invariants that require no additional parallel communication and very
little extra computation to check. This reduces the amount of state
needed to roll back correctly, since we can afford to check these
invariants at every iteration. In fact, GMRES (and variants, like
``Flexible GMRES'') keeps enough state on its own that, unlike in Chen's
work, we do not need to save anything to a persistent store.

Checking invariants naturally fits into the layered approach we
mentioned in the introduction. In the case of FT-GMRES, the outer solver (based
on Flexible GMRES \cite{jje:saad1993flexible}) can check the results of the
unreliable inner solves by computing a residual reliably. The outer
solver will never compute the wrong answer, no matter what the inner
solves do. We present findings in this paper that indicated that a layered
approach coupled with our theory-based detector can tolerate a single SDC event
with little (if any) impact on convergence.
\section{Sandbox Reliability} \label{jje:sec:sandbox}
Relaxing reliability of \emph{all} data and computations may result in
all manner of undesirable and unpredictable behavior.  If
instructions, pointers, array indices, and Boolean values used for
decisions may change arbitrarily at any time, we cannot assert
anything about the results of a computation or the side effects of the
program, even if it runs to completion without abnormal termination.
%
%
%
The least we can do is force the fault-susceptible program to execute
in a \emph{sandbox}.  This is a general idea from computer security,
that allows the execution of untrusted ``guest'' code in a partition
of the computer's state (the ``sandbox'') that protects the rest of
the computer (the ``host'') from the guest's possibly bad behavior.
Sandboxing can even protect the host against malicious code that aims
to corrupt the system's state, so it can certainly handle code subject
to unintentional faults in data and instructions.  

Sandboxes \emph{ensure isolation} of a possibly unreliable phase of
execution.  They \emph{allow data to flow between reliable and
  unreliable} phases of execution.  Also, they let the host
\emph{force guest code to stop} within a predefined finite time, or if
the host suspects the guest may have wandered astray.  This feature is
especially important in distributed-memory computation for preventing
deadlock and other failures due to ``crashed'' or unresponsive nodes.
In general, sandboxing converts some kinds of hard faults into soft
faults, and limits the scope of soft faults to the guest subprogram.


Sandboxing may be implemented in different ways.
For example, the guest may run in a virtual machine on the same
hardware as the host, or the host may be implemented as redundant
processes or systems.  Guests may run on a fast but unreliable
subsystem, and the controlling host program may run on a reliable but
slower subsystem.  We do not specify or depend on a particular
implementation of sandboxing in this paper.


The fault-tolerant inner-outer iteration, described in Section
\ref{jje:sec:ftgmres}, uses the sandbox model.  There, the guest
program performs the task ``Solve a given linear system.''  The host
program invokes the guest repeatedly for different right-hand sides,
and the host performs its own calculations reliably.  Finer-grained
models of reliability may improve the accuracy of the inner solves,
which is what our detector in Section~\ref{jje:sec:gmres:bounds}
accomplishes.

%

The sandbox model of reliability makes only two promises of the
unreliable guest: it returns something (which may not be correct), and
it completes in fixed time.  These already suffice to construct a
working fault-tolerant iterative method, as we will show in Section
\ref{jje:sec:ftgmres}.  However, detecting faults or being able to
limit how faults may occur would also be useful.  
%
%
%
These finer-grained models of reliability can be used to improve
accuracy of the iterative method, or to prove more specific promises
about its convergence.

\section{GMRES} \label{jje:sec:gmres}
The Generalized Minimum Residual method (GMRES) of Saad and Schultz
\cite{jje:saad1986gmres} is a Krylov subspace method for solving
large, sparse, possibly non-symmetric linear systems of the form
$\Mat{A}\Vect{x}=\Vect{b}$.
GMRES is based on the Arnoldi process \cite{mfh:arnoldi1951principle}, which
can also be used to approximate a matrix's eigenvalues and
eigenvectors.  GMRES has the convenient property that the residual
norm of the approximate solution at each iteration is monotonically
non-increasing, assuming correct arithmetic and storage.  Its use of
orthogonal projections and normalized (to length one) basis vectors
also has advantages, that we will discuss below.

We begin this section by explaining how to use properties of the
Arnoldi process to detect faults in an iteration of GMRES.  We then
apply the SDC models we developed above to show how to scale the
linear system in a way that enhances fault detection and bounds the
possible error of the major computational kernels.  We will show in
future work that these bounds by themselves do not suffice to bound
the solution error.  Nevertheless, they can, if one makes inexpensive
changes to how GMRES computes the solution update coefficients.

\subsection{Fault detection via projection coefficients}
\label{jje:sec:gmres:hessenberg}

The norms and inner products that occur in each iteration of the
Arnoldi process in GMRES have a bounded absolute value.  
The bound depends on the norm of the preconditioned matrix, which is
inexpensive to estimate.  We use this bound to detect faults in all the major
computational kernels in GMRES.


\begin{algorithm}
\caption{GMRES}
\label{jje:alg:gmres}
\begin{algorithmic}[1]
\Input{Linear system $\Mat{A}\Vect{x}=\Vect{b}$ and initial guess $\Vect{x}_0$}
\Output{Approximate solution $\Vect{x}_m$ for some $m \geq 0$}
\State{$\Vect{r}_0 := \Vect{b} - \Mat{A} \Vect{x}_0$}\Comment{Unpreconditioned
initial residual vector}
\State{$\beta := \TwoNorm{\Vect{r}_0}$, $\Vect{q}_1 := \Vect{r}_0 / \beta$}
\For{$j = 1, 2, \dots$ until convergence}
\label{jje:alg:gmres:arnoldi_start}
  \State{$\Vect{v}_{j+1} := \Mat{A} \Vect{q}_j$}\Comment{Apply the matrix $A$}
  \label{jje:alg:gmres:orthog_start_vec}
  \For{$i = 1, 2, \dots, j$}\Comment{Orthogonalize}
  \label{jje:alg:gmres:mgs_start}
    \State{$h_{i,j} := \Vect{q}_i \cdot \Vect{v}_{j+1}$}
    \label{jje:alg:gmres:hij}
    \State{$\Vect{v}_{j+1} := \Vect{v}_{j+1} - h_{i,j} \Vect{q}_i $}
    \label{jje:alg:gmres:wi}
  \EndFor
  \label{jje:alg:gmres:mgs_end}
  \State{$h_{j+1,j} := \TwoNorm{\Vect{v}_{j+1}}$}
  \label{jje:alg:gmres:hjp1}
  \If{$h_{j+1,j} \approx 0$} \label{jje:alg:gmres:happyCheck}
       \State{Solution is
       $\Vect{x}_{j-1}$}\label{jje:alg:GMRES:happy_breakdown}\Comment{Happy
       breakdown}
       \State{\textbf{return}}
  \EndIf
  \State{$\Vect{q}_{j+1} := \Vect{v}_{j+1} / h_{j+1,j}$}\Comment{New basis
  vector}
  \label{jje:alg:gmres:arnoldi_end}
  \State{$\Vect{y}_j := \argminy \TwoNorm{ \Mat{H}(\IndexRange{1}{j+1},
  \IndexRange{1}{j}) \Vect{y} - \beta
  \Vect{e}_1}$}
  \State{$\Vect{x}_j := \Vect{x}_0 + [\Vect{q}_1, \Vect{q}_2, \dots, \Vect{q}_j]
  \Vect{y}_j$}\Comment{Compute solution update}
\EndFor
\end{algorithmic}
\end{algorithm}

\subsection{Bounds on the Arnoldi Process}
\label{jje:sec:gmres:bounds}

We start our analysis by bounding the dot product which determines the
$i$-th upper Hessenberg entry, $h_{i,j}$ of the $j$-th Arnoldi
iteration. The Arnoldi process is expressed on
Lines~\ref{jje:alg:gmres:arnoldi_start}--\ref{jje:alg:gmres:arnoldi_end}
in Algorithm~\ref{jje:alg:gmres}.  At its core is an orthogonalization
kernel, which we have chosen to be the Modified Gram-Schmidt (MGS)
process.  Classical Gram-Schmidt or Householder transformations may
also be used.  As we will demonstrate, our bound is invariant of the
orthogonalization algorithm chosen.

The MGS process begins on Line~\ref{jje:alg:gmres:mgs_start} and completes on Line~\ref{jje:alg:gmres:mgs_end}.
To bound $h_{i,j}$ on Line~\ref{jje:alg:gmres:hij}, we exploit a property of
orthogonal projections. It is well known that linear transforms
utilizing orthogonal matrices are \emph{isometric}.  That is, they preserve the
length of the vectors. In $\Rn$, the dot product of a vector with a
unit-length vector is bounded by the length of the first vector.
This means that each $h_{i,j}$ entry is bounded by the length of
the vector that starts the orthogonalization process (the vector we wish to
make orthogonal).
To clarify what ``starts'' the orthogonalization process means, we step back from an algorithmic
formulation, and instead write the orthogonalization kernel as a mathematical
expression.  For clarity, we will use the Classical Gram-Schmidt expression:
\begin{equation}
\Vect{w} = \left[ \Mat{I} - \Mat{Q}^{\text{T}}\Mat{Q} \right]\Vect{u}
\label{jje:eq:classical_gs}
\end{equation}
In Eq~\eqref{jje:eq:classical_gs}, the vector $\Vect{u}$ is what ``starts''
the orthogonalization process, and $\Vect{w}$ is the resulting vector, which is
orthogonal to all vectors in $\Mat{Q}$, where $\Mat{Q} = \{\Vect{q}_1, \dotsc,
\Vect{q}_j\}$.

Returning to Algorithm~\ref{jje:alg:gmres}, what ``starts'' the
orthogonalization process is the vector resulting from
Line~\ref{jje:alg:gmres:orthog_start_vec}. If we can bound the length of this
vector, then we know the maximum absolute value that $h_{i,j}$ can take.
Since we want to bound the length of the resulting vector, we take the
induced $\ell^2$ norm, $\TwoNorm{\Vect{v}_{j+1}} =
\TwoNorm{\Mat{A}\Vect{q}_j}$,
\begin{equation}
\TwoNorm{\Vect{v}_{j+1}} \leq \TwoNorm{\Mat{A}}\TwoNorm{\Vect{q}_j}.
\label{jje:eq:H_bound}
\end{equation}

We can further reduce the bound, recognizing that the basis vector $\Vect{q}_j$
is a unit vector, i.e., $\TwoNorm{\Vect{q}} = 1$. We may deal with
$\TwoNorm{\Mat{A}}$ in several ways: 
\begin{enumerate}
  \item 
$\TwoNorm{\Mat{A}}$ is defined to be
the largest singular value, e.g., $\sigma_\mathrm{max} (\Mat{A})$, or
\item
the 2-norm
is bounded above by the Frobenius norm, which is likely cheaper to compute than
the largest singular value.
\end{enumerate}
 This leads us to an upper bound on \emph{all}
entries in the upper Hessenberg matrix
\begin{align}
|h_{i,j}| \leq \TwoNorm{\Mat{A}} \leq \FNorm{\Mat{A}}.
\label{jje:eq:gmres:bounds:hij}
\end{align}
The bound presented in Eq.~\eqref{jje:eq:gmres:bounds:hij} is crucial, as it
demonstrates that the upper Hessenberg entries are bounded entirely by the input
matrix. In Section~\ref{jje:sec:ftgmres}, we discuss Flexible GMRES
with GMRES (Algorithm~\ref{jje:alg:gmres}) as a preconditioner. In this
scenario, the bound presented is invariant for all applications of the
preconditioner, or, in other words, the bound depends \emph{only} on the input
matrix.
\subsection{Bound Application}

We have shown what the theoretical upper limit is for the values in the upper
Hessenberg. This essentially tells us what is \emph{theoretically possible}
inside the Arnoldi process. Using this approach to construct an
SDC detector is significant. By building a detection scheme in this way, we know
precisely what errors we can detect, and, more importantly, we know what is not
detectable. 

The important factor to keep in mind is that exactly how an error is
committed is irrelevant, the norm bounds allow us to filter out values
that are invalid by theory --- we either detect a
large error or commit a small error, and in Section~\ref{jje:sec:ftgmres} we
will demonstrate how restricting the magnitude of the error committed allows
Flexible GMRES to tolerate the error.

\subsection{Error Detection}

In the context of error detection, we can only detect an error that exceeds the
bound on the upper Hessenberg entry $h_{ij}$. To do this, we insert a
conditional between
Lines~\ref{jje:alg:gmres:hij}~and~\ref{jje:alg:gmres:wi} and 
Lines~\ref{jje:alg:gmres:hjp1}~and~\ref{jje:alg:gmres:happyCheck}
 and test
whether $\Abs{h_{ij}} \leq \FNorm{\Mat{A}}$. Should this condition be invalid, then we assume that we have committed an error 
at some point.

\section{FT-GMRES} \label{jje:sec:ftgmres}
This section describes the Fault-Tolerant GMRES (FT-GMRES) algorithm,
a Krylov subspace method for an iterative solution of large sparse linear
systems of the form $\Mat{A}\Vect{x} = \Vect{b}$.  FT-GMRES computes the correct solution
$\Vect{x}$ even if the system experiences uncorrected faults in both data and
arithmetic~\cite{jje:bridges2012fault}.  
It promises ``eventual convergence'', i.e., it will always either
converge to the right answer, or (in rare cases) stop and report
immediately to the caller if it cannot make progress.  FT-GMRES
accomplishes this by dividing its computations into \emph{reliable}
and \emph{unreliable} phases, using the sandbox model of reliability
described in Section \ref{jje:sec:sandbox}.  Rather than rolling back
any faults that occur in unreliable phases, as a checkpoint / restart
approach would do, FT-GMRES ``rolls forward'' through any faults in
unreliable phases, and uses the reliable phases to drive convergence.
FT-GMRES can also exploit fault detection in order to correct
corrupted data during unreliable phases.

\subsection{FT-GMRES is based on Flexible GMRES}\label{SS:alg:FGMRES}
\begin{algorithm}
\caption{Flexible GMRES (FGMRES)}
\label{jje:alg:FGMRES}
\begin{algorithmic}[1]
\Input{Linear system $Ax=b$ and initial guess $x_0$}
\Output{Approximate solution $x_m$ for some $m \geq 0$}
\State{$\Vect{r}_0 := \Vect{b} - \Mat{A} \Vect{x}_0$}\Comment{Unpreconditioned initial residual}
\State{$\beta := \TwoNorm{\Vect{r}_0}$, $\Vect{q}_1 := \Vect{r}_0 / \beta$}
\For{$j = 1, 2, \dots$ until convergence}
   \State{Solve $\Vect{q}_j = \Mat{M}_j
   \Vect{z}_j$}\label{alg:FGMRES:inner}\Comment{Apply current preconditioner}
  \State{$\Vect{v}_{j+1} := \Mat{A} \Vect{z}_j$}\Comment{Apply the matrix $\Mat{A}$}
  \For{$i = 1, 2, \dots, k$}\Comment{Orthogonalize}
    \State{$h_{i,j} := \Vect{q}_i \cdot \Vect{v}_{j+1}$}
    \State{$\Vect{v}_{j+1} := \Vect{v}_{j+1} - h_{i,j} \Vect{q}_i $}
  \EndFor
  \State{$h_{j+1,j} := \| \Vect{v}_{j+1} \|_2$}
  \State{Update rank-revealing decomposition of $\Mat{H}(\IndexRange{1}{j}, \IndexRange{1}{j})$}
  \If{$\Mat{H}(j+1,j)$ is less than some tolerance}
    \If{$\Mat{H}(\IndexRange{1}{j},\IndexRange{1}{j})$ not full rank}
       \State{Did not converge; report error}\label{alg:FGMRES:rank-def}
     \Else
       \State{Solution is $\Vect{x}_{j-1}$}\Comment{Happy breakdown}
     \EndIf
  \Else
    \State{$\Vect{q}_{j+1} := \Vect{v}_{j+1} / h_{j+1,j}$}
  \EndIf
  \State{$\Vect{y}_j := \argminy \TwoNorm{ \Mat{H}(\IndexRange{1}{j+1},
  \IndexRange{1}{j}) \Vect{y} - \beta
  \Vect{e}_1}$}
  \State{$\Vect{x}_j := \Vect{x}_0 + [\Vect{z}_1, \Vect{z}_2, \dots, \Vect{z}_j]
  \Vect{y}_j$}\label{alg:FGMRES:solution-update}\Comment{Compute solution update}
\EndFor
\end{algorithmic}
\end{algorithm}

FT-GMRES is based on Flexible GMRES (FGMRES)
\cite{jje:saad1993flexible}.  FGMRES extends the Generalized Minimal
Residual (GMRES) method of Saad and Schultz \cite{jje:saad1986gmres}
by ``flexibly'' allowing the preconditioner to change in every
iteration.  An important motivation of flexible methods are
``inner-outer iterations,'' which use an iterative method itself as
the preconditioner (e.g., use GMRES as a preconditioner).  In this
case, ``solve $\Vect{q}_j := \Mat{M}_j \Vect{z}_j$''
Line~\ref{alg:FGMRES:inner} means ``solve the linear system $\Mat{A}
\Vect{z}_j = \Vect{q}_j$ approximately using a given iterative
method.''  For example, suppose GMRES is implemented as a function
$\Vect{x} = \mathsf{gmres}(\Mat{A}, \Vect{b})$, meaning \emph{solve
$\Mat{A}\Vect{x} = \Vect{b}$ for $\Vect{x}$}. Then
Line~\ref{alg:FGMRES:inner} is replaced by $\Vect{z}_j = \mathsf{gmres}(\Mat{A}, \Vect{q}_j)$.

This \emph{inner solve} step preconditions the \emph{outer
solve} (in this case FGMRES).  Changing right-hand sides and possibly
changing stopping criteria for each inner solve means that if one
could express the ``inner solve operator'' as a matrix, it would be
different on each invocation.  This is why inner-outer iterations
require a flexible outer solver.

Flexible methods let the preconditioner change significantly from one
iteration to another; they do not depend on the difference between
successive preconditioners being small.  This is the key observation
behind FT-GMRES: flexible iterations allow successive inner solves to
differ arbitrarily, even unboundedly.  This suggests modeling faulty
inner solves as ``different preconditioners.''  Taking this suggestion
leads to FT-GMRES.

%

There are flexible versions of other iterative methods besides GMRES,
such as CG \cite{jje:golub1999inexact} and QMR
\cite{jje:szyld2001fqmr}, which could also be used as the outer
solver.  We chose FGMRES because it is easy to implement, robust, and
can handle nonsymmetric linear systems.  Experimenting with other
flexible outer iterations is future work.

\subsection{Sandbox Reliability}\label{SS:alg:FT-GMRES}
FT-GMRES further specifies different reliability for inner and outer
solves.  Only inner solves (Line \ref{alg:FGMRES:inner}) are allowed
to run unreliably.  FT-GMRES expects that inner solves do most of the
work, so inner solves run in the less expensive unreliable mode.
Inner solvers need only return with a solution in finite time (see
Section \ref{jje:sec:sandbox}).  That solution may be completely wrong
if errors occurred.

This inner-outer solver approach reduces disruption of existing
solvers.  The outer FGMRES iteration wraps any existing solver with
any preconditioner that it might be using as the inner solver.  Any
solver works, but since we have developed a fault detector in
Section~\ref{jje:sec:gmres}, we chose GMRES as the inner solver.

\subsection{FGMRES' Additional Failure Modes}
\label{jje:sec:fgmres:failure}

FGMRES (and therefore FT-GMRES) have an additional failure mode beyond
those of standard GMRES.  On Line \ref{jje:alg:GMRES:happy_breakdown}
of standard GMRES (Algorithm \ref{jje:alg:gmres}), $ h_{j+1,j} = 0$
indicates that the current iteration produced an invariant subspace.
This means either that we converged to the exact solution, or that the
solve cannot make further progress given the initial guess.  For
FGMRES, if the quantity $h_{j+1,j} = 0$, this does not necessarily
indicate either case.  This is because $\Mat{H}(\IndexRange{1}{j},
\IndexRange{1}{j})$ is always nonsingular in GMRES if $j$ is the
smallest iteration index for which $h_{j+1,j} = 0$, whereas in FGMRES,
$\Mat{H}(\IndexRange{1}{j}, \IndexRange{1}{j})$ may nevertheless be
singular in that case.  (This is Saad's Proposition 2.2
\cite{jje:saad1993flexible}.)  This can happen even in exact
arithmetic.  It may occur due to unlucky choices of the
preconditioners, e.g., $\Mat{M}_j^{-1} = \Mat{A}$ and
$\Mat{M}_{j+1}^{-1} = \Mat{A}^{-1}$.  In practice, this case is rare,
even when inner solves are subject to faults.  Furthermore, it can be
detected inexpensively, since there are algorithms for updating a
rank-revealing decomposition of an $m \times m$ matrix in $O(m^2)$
time (see e.g., Stewart \cite{mfh:stewart1993updating}).  This incurs no
more time than it takes to update the QR factorization of the upper
Hessenberg matrix at iteration $m$.  The ability to detect this rank
deficiency ensures ``trichotomy'' of FGMRES: it either
\begin{enumerate}
\item converges to the desired tolerance,
\item correctly detects an invariant subspace, with a clear indication
  ($h_{j+1,j} = 0$ and $\Mat{H}(\IndexRange{1}{j}, \IndexRange{1}{j})$
  is nonsingular), or
\item gives a clear indication of failure (detected rank deficiency of
  $\Mat{H}(\IndexRange{1}{j}, \IndexRange{1}{j})$).
\end{enumerate}
We base FT-GMRES' ``eventual convergence'' on this trichotomy
property.  In the following section, we will discuss how the
techniques used to detect the third failure case can also be used
to keep the \emph{inner} solves' solutions bounded, as long as faults
in the inner solves are bounded.

\subsection{Fault Tolerance via Regularization}

Both GMRES and Flexible GMRES compute the solution update coefficients
($\Vect{y}_j$ in all algorithms) by solving a small least-squares problem.  This
problem originates from projecting the matrix $\Mat{A}$ onto the Krylov basis,
so we call it the \emph{projected least-squares problem}.  At iteration $k$ (counting
from $k=1$), it has the form
\begin{equation}\label{eq:prj-lsq-problem}
\text{Find $\Vect{y}$ satisfying $\min_y \TwoNorm{ \Mat{H}_k \Vect{y} - \beta
\Vect{e}_1 }$},
\end{equation}
where $\Mat{H}_k$ is a $k+1$ by $k$ upper Hessenberg matrix, $\Vect{y}$
the $k$ coefficients of the solution update, $\beta$ the norm of the
initial residual vector, and $\Vect{e}_1$ the length $k+1$ vector whose first
entry is one and whose remaining entries are zero.  

Saad and Schultz \cite{jje:saad1986gmres} solve this problem by a
structured QR factorization.
%
%
This method lets implementations keep the intermediate reductions of
Steps 1 and 2 at each iteration.  This makes the cost to compute the
solution update $O(k^2)$ coefficients rather than $O(k^3)$.  However,
it can produce unboundedly inaccurate coefficients if the upper
triangular matrix $\Mat{R}_k$ is singular or ill-conditioned.  In
GMRES without faults, this does not normally occur if the matrix
$\Mat{A}$ is not numerically rank deficient.  A numerically
rank-deficient upper Hessenberg matrix normally indicates
convergence\footnote{It may also indicate that the algorithm cannot
  make further progress for the user's choice of initial guess, given
  $\Mat{A}$ and $\Vect{b}$.} at the iteration where it becomes rank
deficient.

%

Linear least-squares problems like \eqref{eq:prj-lsq-problem} always
have a solution.  However, a singular upper Hessenberg matrix may make
the solution set infinite, with unbounded norm.  Unbounded norm in
GMRES' update coefficients means unbounded error in its solution to
$\Mat{A}\Vect{x}=\Vect{b}$.  A \emph{nearly} singular upper Hessenberg matrix
may similarly result in large inaccurate coefficients, by increasing
sensitivity to rounding error in the triangular solve.  In
Section~\ref{jje:sec:fgmres:failure}, we recommended detecting this
case by using a rank-revealing decomposition that supports incremental
updates in order to preserve the $O(k^2)$ cost while detecting rank
deficiency.  This only detects whether the matrix is close to
  singular; it does not prescribe a policy for handling (near)
  singularity.
  
 We define this policy by introducing an additional
constraint, that the solution to \eqref{eq:prj-lsq-problem} have
minimum norm.  We can do this by using a rank-revealing decomposition
that truncates zero singular values.  We can also introduce a
tolerance in order to allow small but nonzero singular values.  This
approach bounds the update coefficients as a function of the largest
singular value of the upper Hessenberg matrix, divided by the least
singular value not truncated.  This is more \emph{robust} than Saad
and Schultz's method, where ``robust'' means ``insensitivity to errors
in the input.''

We can apply the robust technique to the upper triangular system $\Mat{R}_k
\Vect{y} = \Vect{z}_k$, after computing and applying the Givens rotations.  This
is equivalent (in terms of accuracy with respect to rounding error) to a
rank-revealing factorization of $\Mat{H}_k$, and lets us easily
switch ``robustness'' on or off for experiments.  We implemented the
following approaches to solve $\Mat{R}_k \Vect{y} = \Vect{z}_k$:
\begin{enumerate}
\item Standard triangular solve (Saad and Schultz's approach)
\item Attempt a standard triangular solve, and only use a
  rank-revealing method if its solution has \texttt{Inf} or
  \texttt{NaN} values
\item Always use a rank-revealing method
\end{enumerate}
For our experiments, we used a singular-value decomposition as the
rank-revealing factorization, as an easier to implement and no more
accurate substitute for the factorization suggested in our previous
work.  We recommend either Approach 1 or 3.  Approach 2 conceals the
natural error detection that comes with IEEE-754 floating-point data,
without detecting inaccuracy or bounding the error.

\section{Results} \label{jje:sec:results}

To evaluate FT-GMRES and our inner solver bound we explore the impact
on time-to-solution (iteration count) given a fault in \emph{all}
inner solves.

To perform these experiments we developed a two-level solver (``nested
solver'') that uses FT-GMRES as the outer solver, and GMRES as the
inner solver (preconditioner). We used the Trilinos framework
\cite{jje:Trilinos} with FT-GMRES and GMRES implemented as Tpetra
operators.

\subsection{Sample Problems}
We have chosen two sample matrices to demonstrate our technique.  To ensure
reproducibility, we did not create either of these matrices from scratch, rather
we used readily available matrices. The first matrix is fairly common and
arises from the finite difference discretization of the Poisson equation. This
matrix is symmetric and positive definite, meaning that it could be solved using
the Conjugate Gradient method. We generated this matrix
using Matlab's built-in Gallery functionality.
The second matrix chosen presents a more realistic linear system. The
mult\_dcop\_03 matrix comes from the University of Florida Sparse Matrix
Collection \cite{jje:davis2011university}.  It arises from a circuit simulation
problem. The matrix is nonsymmetric and not positive definite, meaning Conjugate
Gradient could not be used to solve the system. The matrix is fairly
small, but is very ill-conditioned, which means that small perturbations may
have a large impact.
We have summarized the characteristics of each matrix in Table~\ref{jje:table:gmres:sample_matrices}.
\begin{table}[htp]\centering
\caption{Sample Matrices}
\label{jje:table:gmres:sample_matrices}
\begin{tabular}{lrr}
\toprule
Properties & Poisson Equation & mult\_dcop\_03 \\
\midrule
number of rows  &
10,000 &
25,187 \\
number of columns  &
10,000 &
25,187 \\
nonzeros  &
49,600 &
193,216 \\
structural full rank?  & 
yes &
yes \\
nonzero pattern symmetry  &
symmetric &
nonsymmetric \\
type  &
real &
real \\
positive definite?  & 
yes &
no \\
Condition Number &
$6.0107 \times 10^{3}$ &
$7.27261 \times 10^{13}$ \\
\midrule
Potential Fault Detectors & & \\
$\TwoNorm{\Mat{A}}$ &
$8$ &
$ 17.1762 $ \\
$\FNorm{\Mat{A}}$ &
$ 446$ &
$ 42.4179$ \\
\bottomrule
\end{tabular}
\end{table}
Note that we have included the potential fault detectors in
Table~\ref{jje:table:gmres:sample_matrices}. These represent the upper bound on
what is acceptable for an upper Hessenberg entry.

\subsubsection{Significance of test problems}
The test problems above represent two classes of matrices: symmetric positive
definite (SPD) and nonsymmetric. Different linear solvers require matrices to have
specific attributes. Conjugate Gradient expects an SPD matrix, while GMRES can
accept both symmetric and nonsymmetric matrices. Relevant to this work, the $\Mat{H}$ matrix
discussed throughout this paper has a unique structure if the
input matrix is symmetric. By structure, we refer to the nonzero pattern of $\Mat{H}$.
For nonsymmetric systems, $\Mat{H}$ is upper Hessenberg, while for SPD systems,
$\Mat{H}$ is tridiagonal (a special case of upper Hessenberg), e.g., see
Figure~\ref{jje:fig:hessenberg_matrices}.
\begin{figure}[H]\centering
\begin{equation*}
\left[\!\!\!
    \begin{array}{cccc}
        \times & \times & \times & \times \\
        \times & \times & \times & \times \\
             0 & \times & \times & \times \\
             0 &      0 & \times & \times \\
    \end{array}
\!\!\!\right]\qquad\text{vs.}\qquad
\left[\!\!\!
    \begin{array}{cccc}
        \times & \times &      0 &  0 \\
        \times & \times & \times &  0 \\
             0 & \times & \times & \times \\
             0 &      0 & \times & \times \\
    \end{array}
\!\!\!\right]
\end{equation*}
\caption{Upper Hessenberg and tridiagonal matrices.}
\label{jje:fig:hessenberg_matrices}
\end{figure}
The fact that solving the Poisson matrix with GMRES \emph{should} create a
tridiagonal matrix is key. This means that specific dot products in the
orthogonalization phase should create entries ``near zero''. If we perturb those
entries (as we are about to do) we can see large penalties in time to solution.

\subsection{Time to Solution Experiments}
In this experiment, we solve a linear system
and determine how many iterations are required to obtain a solution. This is a
\emph{failure-free} run that tells us how many outer iterations (and inner
iterations) are required to obtain a solution.
We then solve the same linear system again (same matrix, right-hand side, and
initial guess), and, on the first iteration of the first inner solve, we perturb
the upper Hessenberg entry ($h_{i,j}$) on the first iteration of the
orthogonalization loop (Line~\ref{jje:alg:gmres:hij} in
Algorithm~\ref{jje:alg:gmres}). We then repeat this process, applying the
fault on all possible inner solve iterations on the first step of the
orthogonalization process.
Note that each experiment injects a single occurrence of SDC.

The choice to inject the fault on the first iteration of the orthogonalization
loop is justified as follows: By faulting early in the orthogonalization phase,
you ``corrupt'' the basis vector from the start, i.e., because we choose
Modified Gram-Schmidt the fault will ``taint'' all subsequent iterations of the
orthogonalization loop (worst-case scenario).

\subsubsection{Fault Values}
To inject a fault, we only need to modify or replace the current $h_{i,j}$ in
Algorithm~\ref{jje:alg:gmres} Line~\ref{jje:alg:gmres:hij} with an incorrect
value.
Directly injecting \texttt{NaN} or \texttt{Inf} reveals nothing, since we can
clearly detect such faults.
We inject an SDC that that represents 3 classes of faults, and these fault values
are relative to the \emph{correct value}:
\begin{enumerate}
  \item very large, $\tilde{h}_{i,j} = h_{i,j} \times 10^{+150}$,
  \item slightly smaller, $\tilde{h}_{i,j} = h_{i,j} \times 10^{-0.5}$, and
  \item very small (nearly zero), $\tilde{h}_{i,j} = h_{i,j} \times 10^{-300}$.
\end{enumerate}
In this experiment, we only record how many iterations it takes to
obtain a solution. It should be noted that for this experiment
parallelism is not a factor, and we are interested in observing how
the solvers behave when perturbed. In particular, this experiment
investigates the solver's behavior when undetectable faults are
injected, and it demonstrates a benefit from filtering obviously
faulty (i.e., large) values. In the following figures, class 2 and 3
faults represent undetectable faults, while class 1 represents a case
that we could detect and to which we could respond, e.g., by halting
the application or restarting the inner solve.

\subsection{Faults in an SPD Problem}
\begin{figure}
    \centering
    \begin{subfigure}[b]{\columnwidth}
		\setlength{\abovecaptionskip}{-0.5\baselineskip}
		\includegraphics[width=\columnwidth]{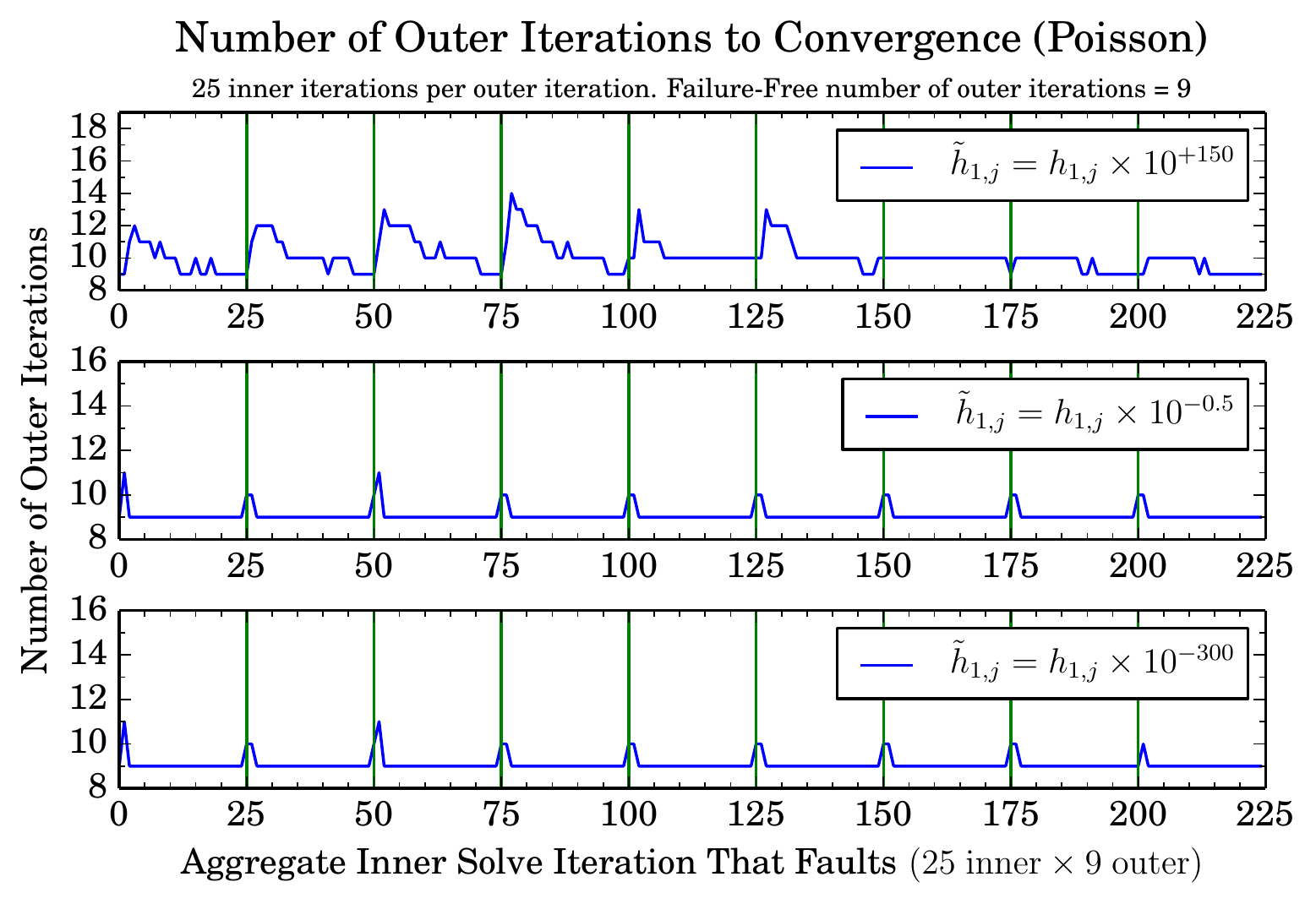}
        \caption{SDC on the first iteration of the Modified
        Gram-Schmidt loop.}
        \label{jje:fig:iteration_plot:poisson100:mgsFirst}
    \end{subfigure}%

    \begin{subfigure}[b]{\columnwidth}
		\setlength{\abovecaptionskip}{-0.5\baselineskip}
		\includegraphics[width=\columnwidth]{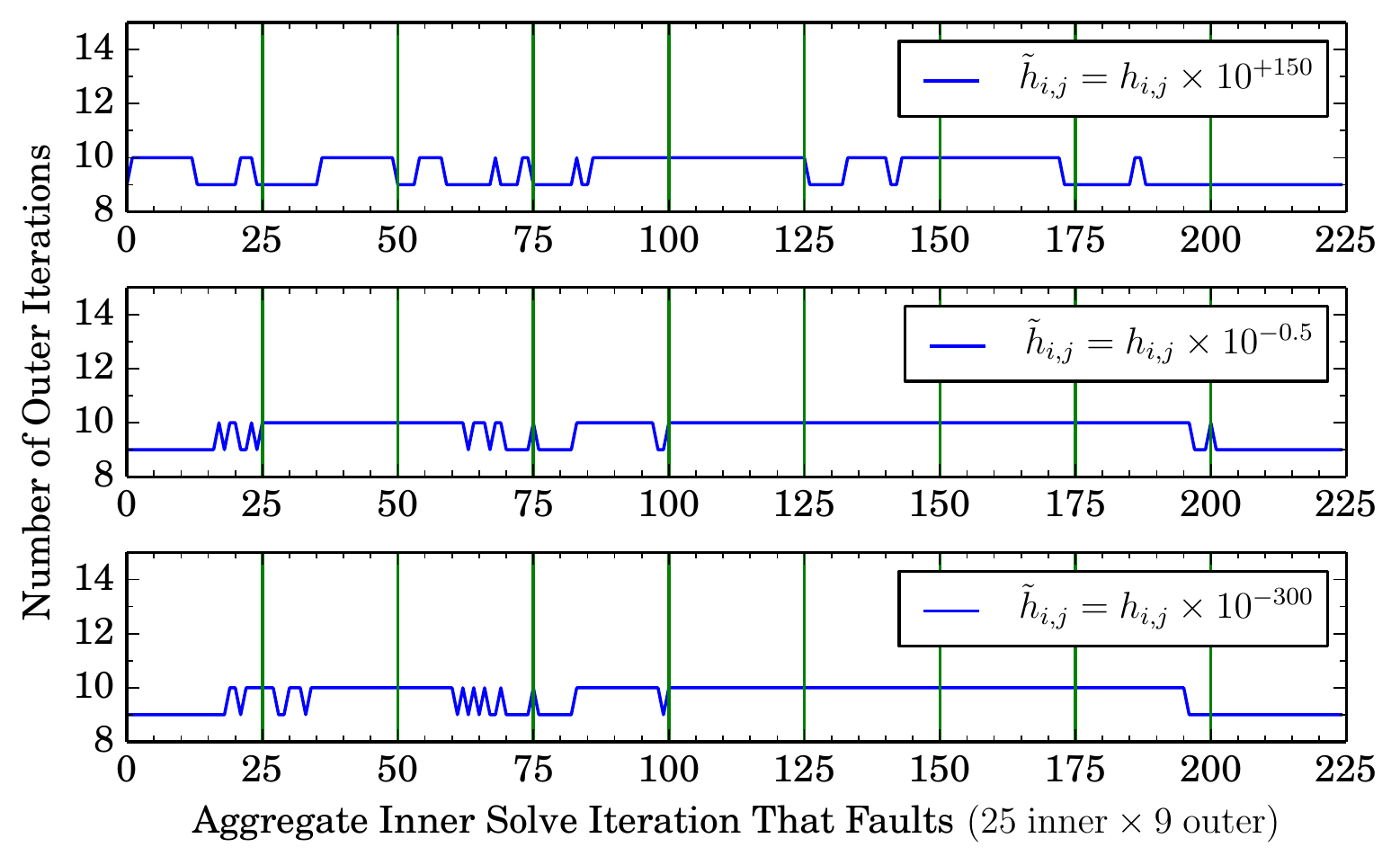}
		\caption{SDC on the last iteration of the Modified
        Gram-Schmidt loop.}
        \label{jje:fig:iteration_plot:poisson100:mgsLast}
    \end{subfigure}%
\setlength{\belowcaptionskip}{-1.5\baselineskip}
\caption{Number of outer iterations required for convergence when solving a
Poisson equation given a single SDC event injected in the orthogonalization
phase of the inner solve. Vertical bars indicate the start of a new inner
solve.}
\label{jje:fig:iteration_plot:poisson100}
\end{figure}

Figure~\ref{jje:fig:iteration_plot:poisson100} illustrates the case of
using GMRES to solve an SPD system of equations. \textbf{In a
failure-free solve FT-GMRES required 9 outer iterations, with each inner solve
performing 25 inner iterations.}
In this case, $\Mat{H}$ should be tridiagonal, meaning that in
Figure~\ref{jje:fig:iteration_plot:poisson100:mgsFirst} for the first
inner solve, the first entry created by the Modified Gram-Schmidt
(MGS) loop, $h_{1,*}$, should be zero from inner iteration 3 onward.
In contrast, Figure~\ref{jje:fig:iteration_plot:poisson100:mgsLast}
faults on the last iteration of the MGS loop, and the last entry in
this column of $\Mat{H}$ can theoretically be nonzero.
\subsubsection{Faulting on the first Modified Gram-Schmidt iteration}
In Figure~\ref{jje:fig:iteration_plot:poisson100:mgsFirst}, we see a
large penalty in time to solution for large faults. This is due to
making entries in $\Mat{H}$ that should be zero, clearly nonzero.
In contrast, if we only slightly perturb these ``near zero'' entries
(class 2 and 3 errors), we see very little impact on time to solution.
The largest increase in outer iterations is two, while the majority of
experiments resulted in no increase in time to solution.
It should be noted that if our fault detector on $h_{i,j}$ was used,
the top plot (large fault) would not be possible.

\subsubsection{Faulting on the last Modified Gram-Schmidt iteration}
Faulting on the last Modified Gram-Schmidt iteration is much different
from faulting on the first for an SPD problem, because the last
$h_{i,j}$ entry created in the orthogonalization phase could
theoretically be nonzero.
From figure~\ref{jje:fig:iteration_plot:poisson100:mgsLast} we see that the worst
case is that we incur one additional outer iteration.
Considering both faults at the start and end of the MGS process, we
see that with our detector we see a maximum increase in outer
iterations to be $2$, in contrast if our detector were not used, we
see increase in outer iterations of $5$.

\subsection{Faulting in a nonsymmetric problem}
We now consider a problem that is not symmetric, meaning that all $h_{i,j}$ we
perturb may be zero, but could also be nonzero --- but each entry in $\Mat{H}$
is still subject to the bound from Eq.~\eqref{jje:eq:gmres:bounds:hij}.
 \textbf{In a
failure-free solve FT-GMRES required 28 outer iterations, with each
inner solve performing 25 inner iterations.}
As in our prior analysis, we consider faults in both the first and last iteration
of the Modified Gram-Schmidt process.

\begin{figure*}
        \centering
        \begin{subfigure}[b]{0.75\textwidth}
			\setlength{\abovecaptionskip}{-0.5\baselineskip}
        	\centering
			\includegraphics[width=\textwidth]{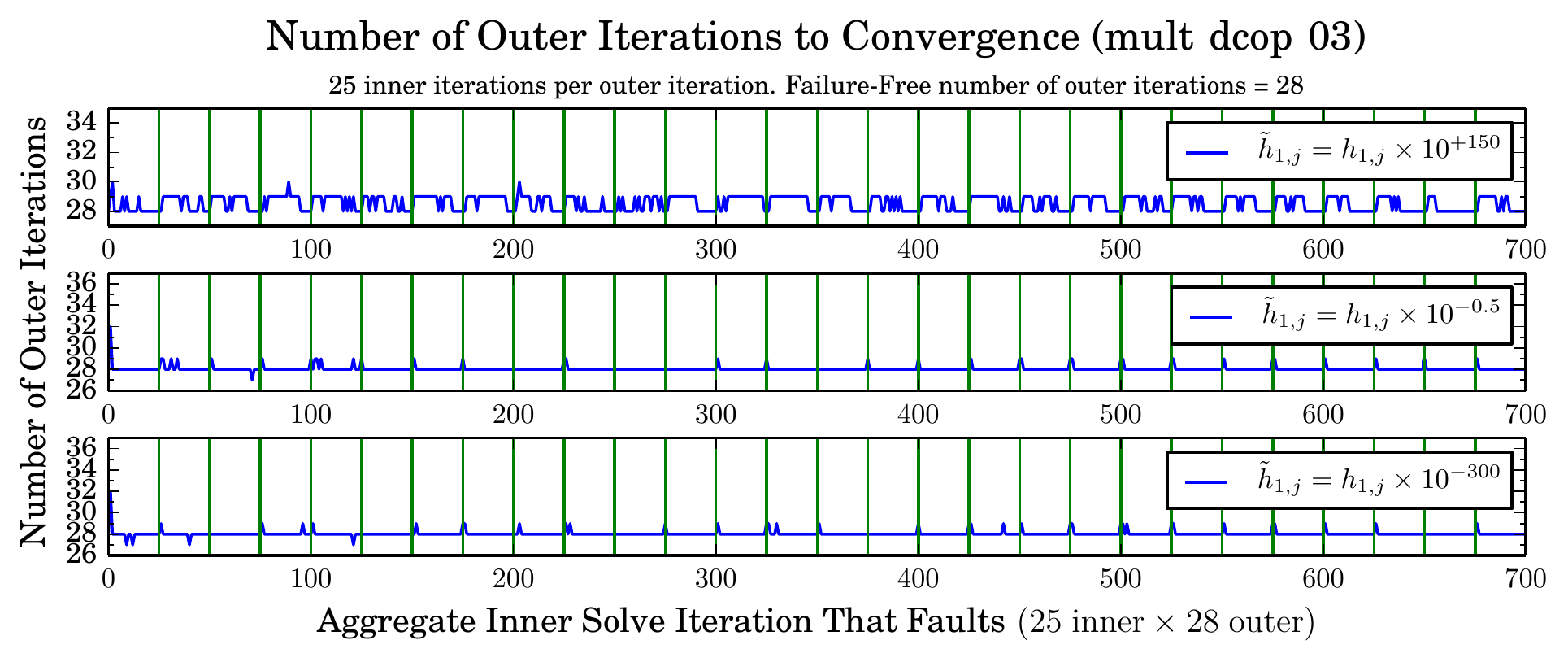}
            \caption{SDC on the first iteration of the Modified
            Gram-Schmidt loop.}
            \label{jje:fig:iteration_plot:mult_dcop_03:mgsFirst}
        \end{subfigure}%

        \begin{subfigure}[b]{0.75\textwidth}
			\setlength{\abovecaptionskip}{-0.5\baselineskip}
        	\centering
			\includegraphics[width=\textwidth]{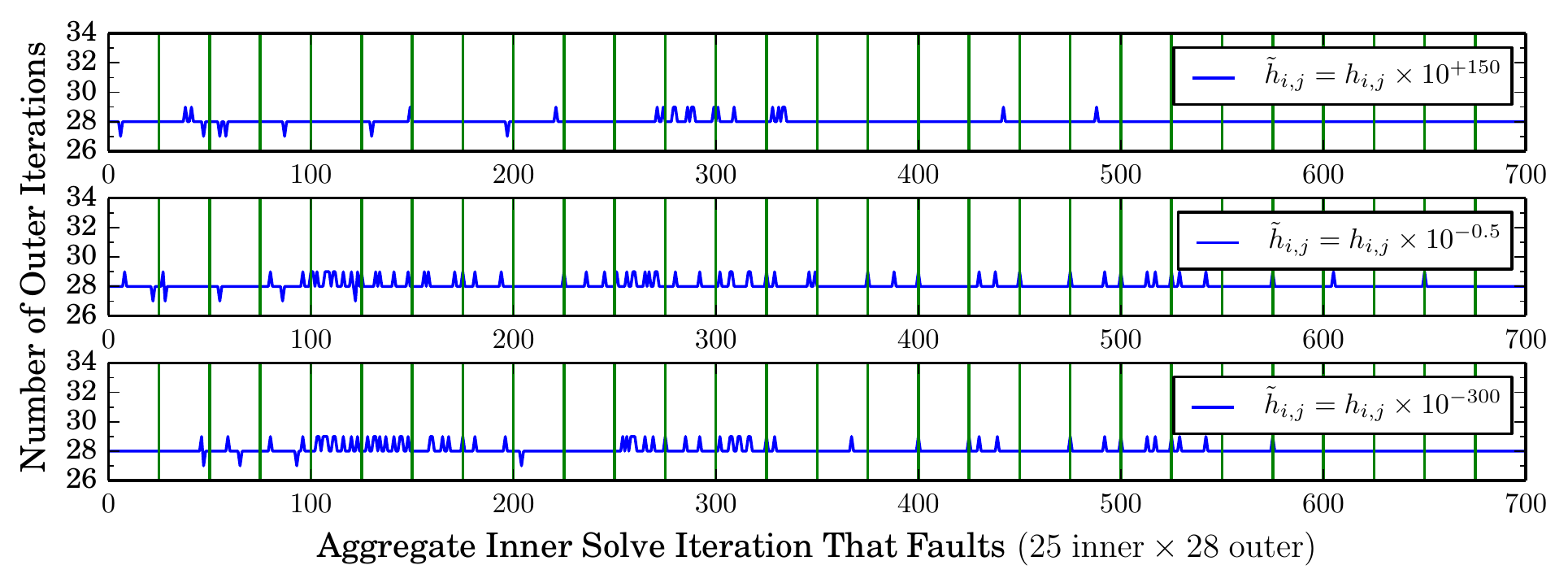}
			\caption{SDC on the last iteration of the Modified
            Gram-Schmidt loop.}
            \label{jje:fig:iteration_plot:mult_dcop_03:mgsLast}
        \end{subfigure}%
\setlength{\belowcaptionskip}{-1.5\baselineskip}
\caption{Number of outer iterations required for convergence when solving the
mult\_dcop\_03 system of equations given a single SDC event injected in the
orthogonalization phase of the inner solve. Vertical bars indicate the
start of a new inner solve.}
\label{jje:fig:iteration_plot:mult_dcop_03}
\end{figure*}

\subsubsection{Faulting on the first Modified Gram-Schmidt iteration}
As expected, in Figure~\ref{jje:fig:iteration_plot:mult_dcop_03} we
see a very different characteristic for faults on the first MGS
iteration. For large faults we see a maximum increase in time to
solution to be 2 outer iterations. For small faults, we see that the
first iteration of the MGS of the first inner solve is extremely
vulnerable to small faults. For class 2 and 3 faults, we see a maximum
increase in outer iterations of 4. If we ignore the first 3 iterations
of the inner solve we see at most 1 additional outer iteration. The
worst-case increase in time to solution actually occurs on the
2\textsuperscript{nd} inner solve iteration, and we leave to future
work's further analysis of the Arnoldi process to explain this
phenomenon. This indicates that additional robustness should be added
at the very start of the first inner solve, and we discuss this briefly in our summary.

\subsubsection{Faulting on the last Modified Gram-Schmidt iteration}
Faulting on the last iteration of the orthogonalization loop again presents a
worst case compared to faulting early. That is, by faulting on the last
orthogonalization iteration, we see an increase in outer iterations in more
cases. We do not the see the sharp increase in iteration count for
faults early in the first inner solve iterations. Note, that the first
MGS iteration is also the last on the first inner solve iteration, and
as stated previously, the first iteration did not exhibit a large
increase in iterations.

\subsection{Summary of Findings}
A common feature seen between both SPD and nonsymmetric solves is that faulting
early in the first inner solves' orthogonalization is universally bad, resulting
in a $33\%$ increase in time to solution for the Poisson problem and $14\%$
increase in time to solution for the mult\_dcop\_03 problem. These
percent increases are not general findings, but we believe this characteristic
will hold true in most, if not all, cases. This may indicate that 
additional effort should be expended early in the first inner solve.
\subsubsection{Performance Characteristics of GMRES}
\label{jje:results:performance_char}
The amount of work per-iteration of GMRES increases linearly. This is
seen in Algorithm~\ref{jje:alg:gmres} in the orthogonalization phase,
where the inner loop iterates from 1 to $j$. Adding redundant
computation early in the inner solve would have minimal performance
impact because the orthogonalization kernel has substantially less
work to perform than in latter iterations. If we included additional
robustness only on the first invocation of the inner solver, we can
mitigate the one edge-case where we see high variability in time to
solution. We leave this to future work.

\subsubsection{Filtering values is cheap and effective}
In all experiments, we find that exploiting the bound on the upper Hessenberg
entries is beneficial, and, in doing so, we typically observe one
additional outer iteration as the penalty should a single SDC event occur. We
believe that this research approach will yield additional invariants that are
cheap to evaluate, and that by combining light-weight mechanisms we drastically
reduce the damage that SDC can introduce.

%

%

\section{Conclusions} \label{jje:sec:conclusion}
In summary, we developed a cheap fault detector for the computational intensive
 orthogonalization stage of GMRES. We then present the FT-GMRES algorithm, and
discuss robustness improvements in the local least squares solve. We then
explained how are detector and robustness modifications can be used to limit the
amount of error that the inner solve may return.

We presented results from two experiments on common classes of matrices that
illustrate that our filtering technique is beneficial, and identified the early
stages of the first inner solve as being the most vulnerable. Furthermore, we
observe that the inner/outer iteration scheme based on FGMRES is extremely
robust to single events of SDC in the orthogonalization phase. We find that this
nested approach, even when not coupled with invariant checks can cope with even
large perturbations introduced by SDC.

\section*{Acknowledgment}\label{S:ack}
This work was supported in part by grants from NSF (awards 1058779 and
0958311) and the U.S.\ Department
of Energy Office of Science, Advanced Scientific Computing Research,
under Program Manager Dr.\ Karen Pao.

Sandia National Laboratories is a multiprogram laboratory managed and
operated by Sandia Corporation, a wholly owned subsidiary of Lockheed
Martin Corporation, for the U.S.\ Department of Energy's National
Nuclear Security Administration under contract DE-AC04-94AL85000.

\bibliographystyle{IEEE}
\bibliography{paper}

\begin{thebibliography}{10}
\providecommand{\url}[1]{#1}
\csname url@samestyle\endcsname
\providecommand{\newblock}{\relax}
\providecommand{\bibinfo}[2]{#2}
\providecommand{\BIBentrySTDinterwordspacing}{\spaceskip=0pt\relax}
\providecommand{\BIBentryALTinterwordstretchfactor}{4}
\providecommand{\BIBentryALTinterwordspacing}{\spaceskip=\fontdimen2\font plus
\BIBentryALTinterwordstretchfactor\fontdimen3\font minus
  \fontdimen4\font\relax}
\providecommand{\BIBforeignlanguage}[2]{{%
\expandafter\ifx\csname l@#1\endcsname\relax
\typeout{** WARNING: IEEEtran.bst: No hyphenation pattern has been}%
\typeout{** loaded for the language `#1'. Using the pattern for}%
\typeout{** the default language instead.}%
\else
\language=\csname l@#1\endcsname
\fi
#2}}
\providecommand{\BIBdecl}{\relax}
\BIBdecl

\bibitem{jje:bridges2012fault}
P.~G. {Bridges}, K.~B. {Ferreira}, M.~A. {Heroux}, and M.~{Hoemmen},
  ``{Fault-tolerant linear solvers via selective reliability},'' \emph{ArXiv
  e-prints}, Jun. 2012.

\bibitem{jje:intel:fdiv}
Intel, ``{FDIV} replacement program: Description of the flaw,'' Jul. 2004.

\bibitem{jje:saad1993flexible}
Y.~Saad, ``A flexible inner-outer preconditioned {GMRES} algorithm,''
  \emph{SIAM J. Sci. Comput.}, vol.~14, no.~2, pp. 461--469, Mar. 1993.

\bibitem{jje:Trilinos}
M.~A. Heroux \emph{et~al.}, ``An overview of the {Trilinos} project,''
  \emph{ACM Trans. Math. Softw.}, vol.~31, no.~3, pp. 397--423, 2005.

\bibitem{jje:asanovic2006landscape}
K.~Asanovic, R.~Bodik, B.~C. Catanzaro, J.~J. Gebis, P.~Husbands, K.~Keutzer,
  D.~A. Patterson, W.~L. Plishker, J.~Shalf, S.~W. Williams, and K.~A. Yelick,
  ``The {L}andscape of {P}arallel {C}omputing {R}esearch: {A} {V}iew from
  {B}erkeley,'' EECS Department, University of California, Berkeley, Tech. Rep.
  UCB/EECS-2006-183, Dec 2006.

\bibitem{jje:asanovic2009landscape}
K.~Asanovic, R.~Bodik, J.~W. Demmel, T.~Keaveny, K.~Keutzer, J.~Kubiatowicz,
  N.~Morgan, D.~A. Patterson, K.~Sen, J.~Wawrzynek, D.~Wessel, and K.~A.
  Yelick, ``A {V}iew of the {P}arallel {C}omputing {L}andscape,''
  \emph{Communications of the ACM}, vol.~52, no.~10, pp. 56--67, 2009.

\bibitem{jje:kogge2008exascale}
P.~M. Kogge \emph{et~al.}, ``{ExaScale Computing Study: Technology Challenges
  in Achieving Exascale Systems},'' University of Notre Dame CSE Department,
  Tech. Rep. TR-2008-13, September 2008.

\bibitem{jje:karnik2004characterization}
T.~Karnik, P.~Hazucha, and J.~Patel, ``Characterization of soft errors caused
  by single event upsets in {CMOS} processes,'' \emph{IEEE Trans. Dependable
  Secur. Comput.}, vol.~1, pp. 128--143, April 2004.

\bibitem{jje:miskov-zivanov2007soft}
N.~Miskov-Zivanov and D.~Marculescu, ``Soft error rate analysis for sequential
  circuits,'' in \emph{Proceedings of the Conference on Design, Automation and
  Test in Europe}, ser. DATE '07.\hskip 1em plus 0.5em minus 0.4em\relax San
  Jose, CA, USA: EDA Consortium, 2007, pp. 1436--1441.

\bibitem{jje:haque2010hard}
I.~S. Haque and V.~S. Pande, ``Hard data on soft errors: A large-scale
  assessment of real-world error rates in {GPGPU},'' in \emph{Proceedings of
  the 2010 10th IEEE/ACM International Conference on Cluster, Cloud and Grid
  Computing}, ser. CCGRID '10.\hskip 1em plus 0.5em minus 0.4em\relax
  Washington, DC, USA: IEEE Computer Society, 2010, pp. 691--696.

\bibitem{lammers2010era}
D.~Lammers, ``The era of error-tolerant computing,'' \emph{IEEE Spectr.},
  vol.~47, no.~11, p.~15, Nov. 2010.

\bibitem{jje:iterative:Shantharam:2011}
M.~Shantharam, S.~Srinivasmurthy, and P.~Raghavan, ``Characterizing the impact
  of soft errors on iterative methods in scientific computing,'' in
  \emph{Proceedings of the international conference on Supercomputing}, ser.
  ICS '11.\hskip 1em plus 0.5em minus 0.4em\relax New York, NY, USA: ACM, 2011,
  pp. 152--161.

\bibitem{jje:iterative:Shantharam:2012}
------, ``Fault tolerant preconditioned conjugate gradient for sparse linear
  system solution,'' in \emph{Proceedings of the 26th ACM international
  conference on Supercomputing}, ser. ICS '12.\hskip 1em plus 0.5em minus
  0.4em\relax New York, NY, USA: ACM, 2012, pp. 69--78.

\bibitem{jje:iterative:Sloan:2012}
J.~Sloan, R.~Kumar, and G.~Bronevetsky, ``Algorithmic approaches to low
  overhead fault detection for sparse linear algebra,'' in \emph{Proceedings of
  the 2012 42nd Annual IEEE/IFIP International Conference on Dependable Systems
  and Networks (DSN)}, ser. DSN '12.\hskip 1em plus 0.5em minus 0.4em\relax
  Washington, DC, USA: IEEE Computer Society, 2012, pp. 1--12.

\bibitem{jje:iterative:Bronevetsky:2008}
G.~Bronevetsky and B.~de~Supinski, ``Soft error vulnerability of iterative
  linear algebra methods,'' in \emph{Proceedings of the 22nd annual
  international conference on Supercomputing}, ser. ICS '08.\hskip 1em plus
  0.5em minus 0.4em\relax New York, NY, USA: ACM, 2008, pp. 155--164.

\bibitem{jje:Michalak:2012}
S.~Michalak, A.~Dubois, C.~Storlie, H.~Quinn, W.~Rust, D.~DuBois, D.~Modl,
  A.~Manuzzato, and S.~Blanchard, ``Assessment of the impact of
  cosmic-ray-induced neutrons on hardware in the {R}oadrunner supercomputer,''
  \emph{Device and Materials Reliability, IEEE Transactions on}, vol.~12,
  no.~2, pp. 445--454, 2012.

\bibitem{jje:Mueller2013preprint}
E.~{Mueller} and R.~{Scheichl}, ``Massively parallel solvers for elliptic
  {PDEs} in numerical weather and climate prediction,'' \emph{ArXiv e-prints},
  Jun. 2013.

\bibitem{jje:chen2013online}
Z.~Chen, ``{Online-ABFT}: an online algorithm based fault tolerance scheme for
  soft error detection in iterative methods,'' in \emph{Proceedings of the 18th
  ACM SIGPLAN symposium on Principles and practice of parallel programming},
  ser. PPoPP '13.\hskip 1em plus 0.5em minus 0.4em\relax New York, NY, USA:
  ACM, 2013, pp. 167--176.

\bibitem{jje:saad1986gmres}
Y.~Saad and M.~H. Schultz, ``{GMRES}: A generalized minimal residual algorithm
  for solving nonsymmetric linear systems,'' \emph{SIAM J. Sci. Stat. Comput.},
  vol.~7, no.~3, pp. 856--869, Jul. 1986.

\bibitem{mfh:arnoldi1951principle}
W.~E. Arnoldi, ``The principle of minimized iterations in the solution of the
  matrix eigenvalue problem,'' \emph{Quarterly of Applied Mathematics}, vol.~9,
  pp. 17--29, 1951.

\bibitem{jje:golub1999inexact}
G.~H. Golub and Q.~Ye, ``Inexact preconditioned conjugate gradient method with
  inner-outer iteration,'' \emph{SIAM J. Sci. Comput.}, vol.~21, pp.
  1305--1320, 1999.

\bibitem{jje:szyld2001fqmr}
D.~B. Szyld and J.~A. Vogel, ``{FQMR}: A flexible quasi-minimal residual method
  with inexact preconditioning,'' \emph{SIAM J. Sci. Comput.}, vol.~23, no.~2,
  pp. 363--380, 2001.

\bibitem{mfh:stewart1993updating}
G.~W. Stewart, ``Updating a rank-revealing {$ULV$} decomposition,'' \emph{SIAM
  J. Matrix Anal. Appl.}, vol.~14, no.~2, pp. 494--499, April 1993.

\bibitem{jje:davis2011university}
T.~A. Davis and Y.~Hu, ``The {U}niversity of {F}lorida {S}parse {M}atrix
  {C}ollection,'' \emph{ACM Transactions on Mathematical Software}, vol.~38,
  no.~1, pp. 1:1--1:25, 2011.

\end{thebibliography}

\end{document}